\documentclass[11pt]{article}

\pdfoutput=1
\usepackage{graphicx}
\usepackage{cancel}
\usepackage{amssymb}
\usepackage{amsmath,dsfont,textcomp}
\usepackage{bm}
\usepackage{color}
\usepackage{euscript,amsmath}
\usepackage{epsf}
\usepackage{fancyhdr}
\usepackage{cite}

\usepackage{subfigure}
\usepackage{slashed}
\usepackage{simplewick}
\usepackage[usenames,dvipsnames]{xcolor}
\usepackage{rotating}

\def\lsim{\mathrel{\rlap{\lower4pt\hbox{\hskip1pt$\sim$}}
    \raise1pt\hbox{$<$}}}         
\def\gsim{\mathrel{\rlap{\lower4pt\hbox{\hskip1pt$\sim$}}
    \raise1pt\hbox{$>$}}}

\def\beq{\begin{equation}}
\def\eeq{\end{equation}}
\def\bea{\begin{eqnarray}}
\def\eea{\end{eqnarray}}
\def\bitem{\begin{itemize}}
\def\eitem{\end{itemize}}
\def\ba{\begin{array}}
\def\ea{\end{array}}
\def\bal{\begin{align}}
\def\eal{\end{align}}
\def\bi{\begin{itemize}}
\def\ei{\end{itemize}}
\def\lsim{\mathrel{\rlap{\lower4pt\hbox{\hskip1pt$\sim$}}
    \raise1pt\hbox{$<$}}}         
\def\gsim{\mathrel{\rlap{\lower4pt\hbox{\hskip1pt$\sim$}}
    \raise1pt\hbox{$>$}}}

\newcommand{\newc}{\newcommand}
\newc{\renewc}{\renewcommand}

\newc{\ie}{{\it i.e.~}}          \newc{\etal}{{\it et al.~}}
\newc{\eg}{{\it e.g.~}}          \newc{\etc}{{\it etc.~}}
\newc{\cf}{{\it c.f.~}}
\newc{\os}{\mbox{\hspace{4pt}}}
\newc{\us}{\mbox{\hspace{12pt}}}

\renewc{\bar}{\overline}

\newc{\gev}{\,{\rm GeV}}
\newc{\mev}{\,{\rm MeV}}
\newc{\ev}{\,{\rm eV}}
\newc{\kev}{\,{\rm keV}}
\newc{\tev}{\,{\rm TeV}}
\def\ln{\mathop{\rm ln}}

\newc{\LM}{\mathcal{L}}
\newc{\SM}{\mathcal{S}}

\newc{\HM}{\mathcal{H}}
\newc{\GM}{\mathcal{G}}
\newc{\OM}{\mathcal{O}}
\newc{\FM}{\mathcal{F}}
\newc{\AM}{\mathcal{A}}
\newc{\BM}{\mathcal{B}}
\newc{\NM}{\mathcal{N}}
\newc{\WM}{\mathcal{W}}
\newc{\ZM}{\mathcal{Z}}

\newc{\Chi}{\mathcal{X}}

\newcommand{\be}{\begin{equation}}
\newcommand{\ee}{\end{equation}}

\newcommand{\brH}{$\text{BR}_{H}$}
\newcommand{\brL}{$\text{BR}_{L}$}
                           
\begin{document}

\twocolumn[ 

\title{\bf\Large AMS-02 and Next-to-Minimal Universal Extra Dimensions}


\author{Yu Gao$^{1}$, Kyoungchul Kong$^2$, and Danny Marfatia$^{3}$\\[2ex]
\small\it $^{1}$Mitchell Institute for Fundamental Physics and Astronomy,\\ 
\small\it Department of Physics and Astronomy, Texas A\&M University, College Station, TX 77843, U.S.A.\\
\small\it $^{2}$Department of Physics and Astronomy, University of Kansas, Lawrence, KS 66045, U.S.A.\\
\small\it $^3$Department of Physics and Astronomy, University of Hawaii at Manoa, Honolulu, HI 96822, U.S.A}
\date{}
\maketitle

\vspace{-1.1cm}

\begin{quote}
The anomaly detected by AMS-02 and PAMELA in the cosmic-ray positron flux when interpreted as arising from dark matter annihilation suggests that dark matter may interact differently with hadrons and leptons so as to remain compatible with cosmic-ray antiproton data.
Such a scenario is readily accommodated in models with extra spatial dimensions. We study indirect detection of Kaluza-Klein (KK) dark matter in Universal Extra Dimensions with brane-localized terms and 
fermion bulk masses: Next-to-Minimal Universal Extra Dimensions. So that an excess of antiprotons is not produced in explaining the positron anomaly, it is necessary that the KK bulk masses in the lepton and hadron sectors be distinct. Even so, we find that cosmic-ray data disfavor a heavy KK photon dark matter scenario. Also, we find these scenarios with flavor-universal bulk masses to be in conflict with dijet and dilepton searches at the LHC.
\end{quote}
]


While physics beyond the standard model (SM) often involves extensions to gauge symmetry groups, models with extra spatial dimensions have attracted great interest recently. Models with Universal Extra Dimensions (UED)~\cite{Appelquist:2000nn} provide a useful framework that yields a degenerate mass spectrum 
of new particles with identical spins to their SM partners.
In such models, Kaluza-Klein (KK) bosons~\cite{servanttait} and KK neutrinos are good dark matter (DM) candidates, 
with the KK photon extensively studied because of two noteworthy features:
its non-relativistic self-annihilation does not suffer from helicity suppression, and leptonic final states are preferred; for a recent review see Ref.~\cite{review}. 
 
The simplest models with UED are defined in five dimensions with $S_1/Z_2$. 
The KK spectrum is entirely fixed by the renormalization group running between the ultraviolet cutoff and the electroweak scale, and the
additional assumption of vanishing boundary conditions at the cutoff scale. 
This so-called Minimal UED (MUED) model has only two parameters, $R^{-1}$ and $\Lambda$, the compactification and cutoff scales, respectively~\cite{Cheng}.

A very different mass spectrum from that of MUED is obtained on the inclusion of brane-localized terms or fermion-bulk masses, thus broadening the implications for collider and astrophysical phenomenology. 
Brane-localized terms are generated by quantum corrections even if they are absent at tree level. 
The boundary terms in such non-minimal UED models lower the masses of both KK fermions and bosons~\cite{FMP}.
In models with fermion bulk masses but no boundary terms, known as Split UED, KK fermion masses are enhanced, akin to split supersymmetry~\cite{sUED1,sUED2,Huang:2012kz}. 
Recently, both boundary and bulk terms have been considered together in Ref.~\cite{Flacke:2013pla}. We refer to such models as Next-to-Minimal UED or NMUED, and it is in their context that we study recent cosmic ray anomalies.

Annihilation of KK DM in the Milky Way halo can potentially modify local cosmic ray fluxes and cause
the anomalies seen  by the PAMELA~\cite{Adriani:2008zr} and AMS-02~\cite{Aguilar:2013qda} experiments.
The induced $e^{\pm}$ and antiproton fluxes can be compared with AMS-02~\cite{Aguilar:2013qda} and PAMELA~\cite{Adriani:2010rc} data, respectively. 
In this Letter, we study if  NMUED offers a satisfactory explanation of the PAMELA and AMS-02 positron anomaly.

\label{sect:relic}

KK masses and couplings are modified significantly in the presence of bulk masses ($\mu$) and brane localized terms (with a coefficient $r$), 
which in turn affect dark matter annihilation.  
For $\mu<0$,
KK fermion masses increase with $|\mu|$,
while both KK fermion and boson masses decrease as the brane parameter $r$ is increased. 
On the other hand, for $\mu>0$, the KK photon is heavier than KK fermions and is not a viable dark matter candidate. 
For this reason, we restrict ourselves to $\mu < 0$.

In the universal parametrization (where the same $\mu$ and $r$ are shared by all KK fermions) branching fractions of KK photon annihilation into SM particles are 
0.2 for each charged lepton, 0.035 for the three neutrino families, 0.11 for the top pair, 0.0073 for the bottom pair and 0.25 for light quarks.
Higgs final states are negligible. Equity in the leptonic and hadronic terms leads to constant relative ratios of annihilation branching fractions into SM particles. In the universal case the hadronic channels make up 35\% of the total annihilation rate, making it impossible to obtain consistency with PAMELA's antiproton data.  In what follows, we allow the bulk masses and brane terms to be different for the KK lepton ($\mu_L$ and $r_L$) and quark ($\mu_Q$ and $r_Q$) sectors, but require that they be flavor blind within each sector.

%


%

\label{sect:astro}


With flavor-blind $\mu$ and $r$, KK photon annihilation has equal branching fractions into the three charged leptons. The relative branching fractions of the hadronic channels are $q\bar{q}:t\bar{t}:b\bar{b}=1:0.47:0.03$, where $q$ denotes the light quarks. We define \brH$=\text{BR}_{b\bar{b}}+\text{BR}_{t \bar{t}}+ \text{BR}_{ q \bar{q}}$ and \brL$=\sum_{l}\text{BR}_{l^+l^-}$ to be the total hadronic and leptonic branching fractions. In NMUED, {\brH} + {\brL}$=1$ since annihilation into gauge bosons, Higgs bosons and neutrinos is negligible.

We analyze the separate $e^-$ and $e^+$ energy spectra from AMS-02 which has more information than the $e^+$ fraction. The KK photon mass appears as the end point of the $e^\pm$ spectrum.   With AMS-02 data alone, the KK photon is not required to be heavier than 1~TeV. Note that our data analysis does not depend on the details of NMUED.

\begin{figure}[t]
\includegraphics[width=0.4\textwidth]{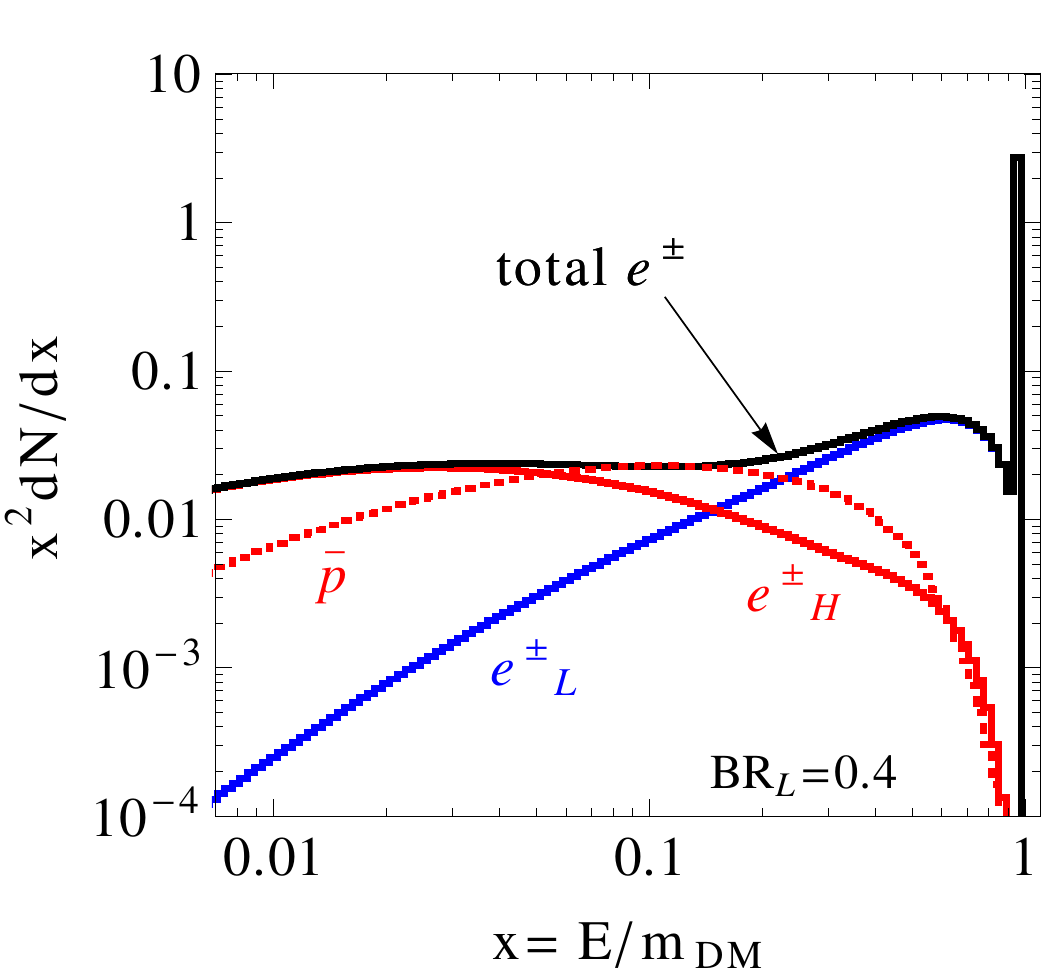}
\caption{$e^\pm$ (solid) and antiproton (dotted) injection spectra from leptonic (blue) and hadronic (red) annihilation channels, with \brL=0.4 and \brH=0.6. The end-point peak is due to DM annihilation into $e^+e^-$. The DM mass is 1 TeV.}
\label{fig:prompt}
\end{figure}

The injection spectra from DM annihilation are calculated with {\tt MadGraph/MadEvent} \cite{Alwall:2007st} with showering implemented with {\tt Pythia}~\cite{Sjostrand:2006za}.  We adopt an 
Einasto~\cite{Navarro:2008kc} profile for the dark matter density distribution in the galactic halo. The propagation to the Earth is computed using {\tt Galprop}~\cite{Strong:1999sv}.

In Fig.~\ref{fig:prompt} we show the $e^\pm$ injection spectra from the leptonic and hadronic channels  with 
\brH=0.6 and \brL=0.4, which are the best-fit values for a 1~TeV KK photon. The hadronic channels soften the $e^-$ and $e^+$ spectra in the low energy region and lend better agreement with AMS-02 data. 
In order to accommodate the AMS-02 $e^-$ data, it is necessary to vary the astrophysical $e^-$ background above 10~GeV; we assume the background is well-described by an unbroken power-law in the energy range relevant to AMS-02. We calculate the galactic background and DM signal on a five-parameter grid. We vary the normalization and spectral index for astrophysical electrons, and vary three parameters that describe particle diffusion in the galactic magnetic field. For details see Ref.~\cite{Barger:2009yt}. 

\begin{figure*}[t]
\centerline{ 
\includegraphics[width=0.3\textwidth]{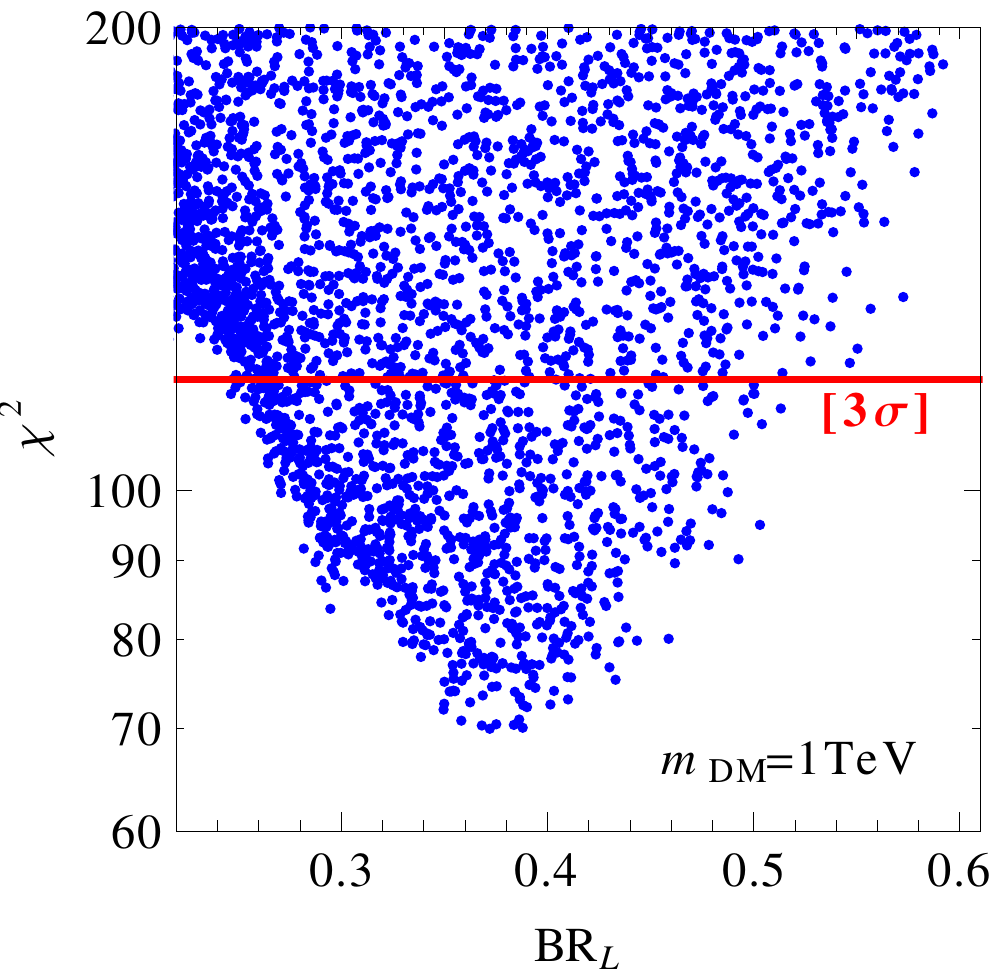}
\includegraphics[width=0.307\textwidth]{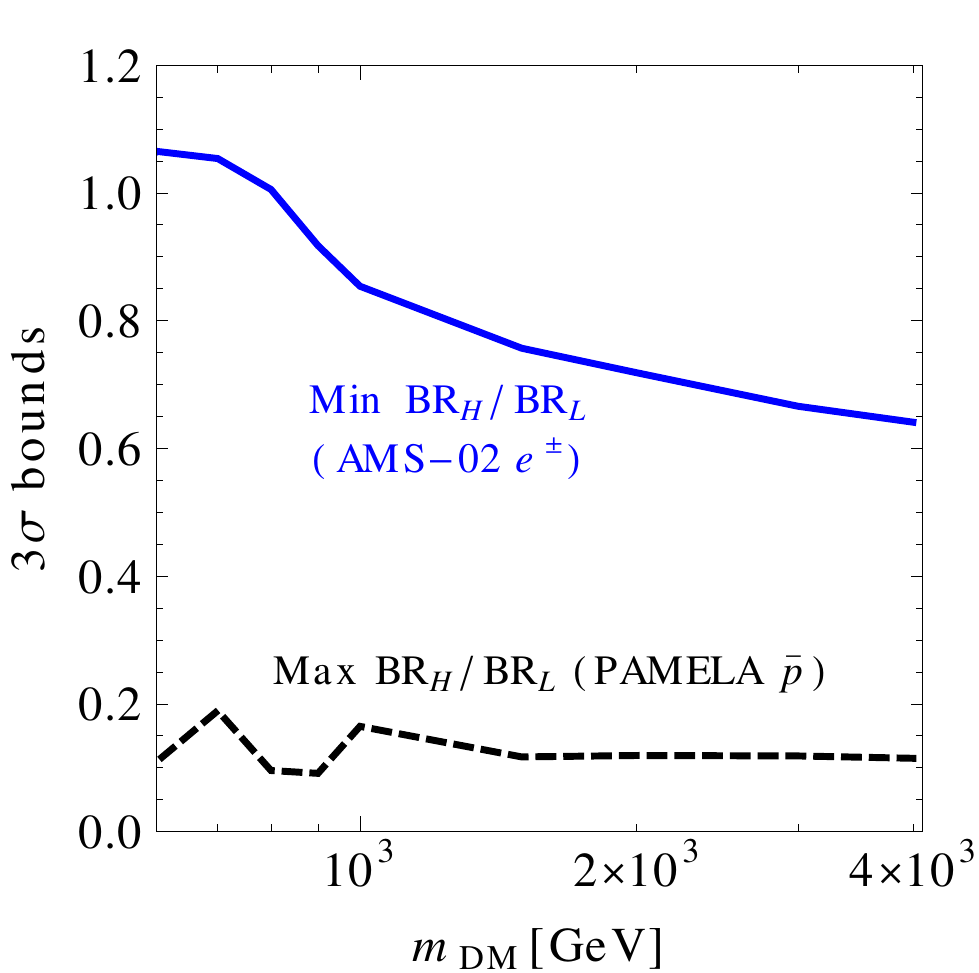}
\includegraphics[width=0.31\textwidth]{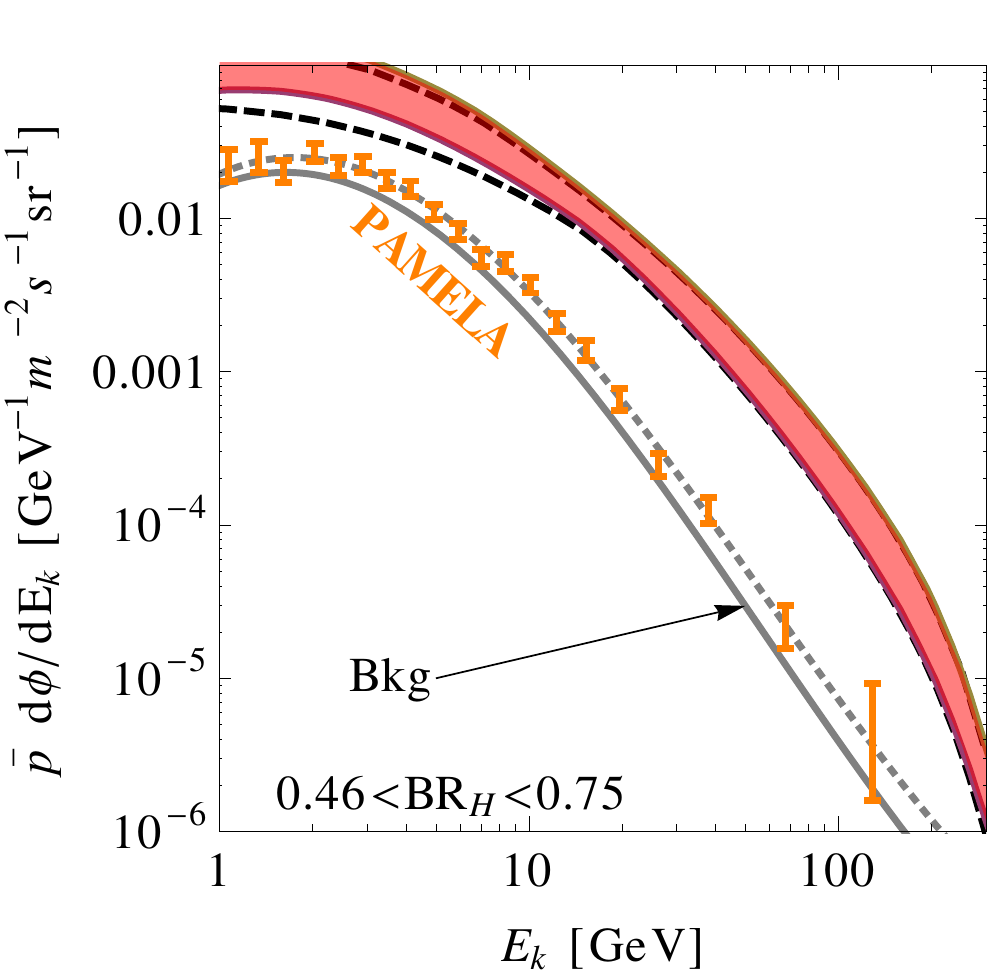}
}
\caption{{\it Left:} Fits to AMS-02 $e^-$ and $e^+$ spectra for different leptonic branching fractions, for a 1~TeV KK photon. Variations in the galactic $e^\pm$ background and boost factor are marginalized over for each point.
{\it Middle:} $3\sigma$ bounds on $\text{BR}_{H}/\text{BR}_{L}$ from the AMS-02 $e^\pm$ spectra and the PAMELA antiproton spectrum. {\it Right:} The shaded region shows the $\bar{p}$ signal+background spectra for $m_{DM}=1$~TeV that correspond to the 3$\sigma$ fit to AMS-02.  The DM-only contributions (black dashed) are higher than the PAMELA data, even without considering the galactic background. The separation between the polynomial fit to the $\bar{p}$ data (gray dotted) and an extreme possibility for the background (labelled ``Bkg") gives an idea of the size of the deviation permitted by our modeling of the background uncertainty. 
}
\label{fig:fitresults}
\end{figure*}

In our fit to AMS-02 $e^\pm$ data above 10~GeV, we allow $\text{BR}_{L}$ to vary between 0 and 1. We find an ensemble of fits at the $3\sigma$ confidence level by requiring that the reduced $\chi^2$ be smaller than 1.5 for 78 degrees of freedom. The left panel of Fig.~\ref{fig:fitresults} shows a sample set of fits for $m_{DM}=$1~TeV. Note that the hadronic channels make up at least 46\% of the total branching within $3\sigma$. This is because the high precision AMS-02 data require soft low-energy $e^\pm$ spectra. As a result, a significant addition of shower-produced $e^\pm$ is necessary, as shown in Fig.~\ref{fig:prompt}, to sufficiently soften the hard $e^\pm$ spectra from leptonic annihilation channels.

Since the leptonic channels determine the high energy part of the $e^\pm$ signal spectra, the quantity BF$\cdot \left<v\sigma\right> \times \text{BR}_{L}$ (where BF is the usual factor that boosts annihilations) is fixed by data and the galactic background. 
In analogy, only hadronic channels produce antiprotons, so BF$\cdot \left<v\sigma\right> \times \text{BR}_{H}$ alone determines the fit to antiproton data.  Thus, constraints on $\text{BR}_{H}/\text{BR}_{L}$ can be placed using the $e^\pm$ and $\bar{p}$ data. For each point in the $3\sigma$-fit region to AMS-02, we calculate the DM induced $\bar{p}$ signal and fit to PAMELA's $\bar{p}$ data. We assume there to be no excess in the antiproton data and  parameterize the galactic antiproton background by 
\be 
\left.\frac{d\phi}{dE_k}\right|_{bkg}={C}\left(\frac{E_k}{E_0}\right)^{\delta}e^{-3.94 + 0.91x - 0.86 x^2 + 0.054 x^3 + 
 0.0021 x^4},
\ee
where $E_k$ is the antiproton's kinetic energy in GeV and $x\equiv \ln\left(\frac{E_k}{\text{GeV}}\right)$. Here $C=1$~GeV$^{-1}$m$^{-2}$s$^{-1}$sr$^{-1}$, $E_0=30$~GeV and $\delta=0$. As a conservative estimate of the uncertainty in the $\bar{p}$ background, we allow the overall normalization to vary between $0.6C$ and $1.4C$, and the spectral index $\delta$ to vary between $-0.1$ and +0.1~\cite{apbkg}. For consistency at $3\sigma$, we require the combined DM and background flux contribution to fit the antiproton data with
$\Delta\chi^2\le 9$ compared to that with only the background.  The maximum allowed value of $\text{BR}_{H}/\text{BR}_{L}$ at $3\sigma$ is found to be less than about 0.2, as illustrated by the dashed black curve in the middle panel of Fig.~\ref{fig:fitresults}. In comparison, the minimum value of $\text{BR}_{H}/\text{BR}_{L}$ to fit 
the AMS-02 $e^\pm$ data at 3$\sigma$ is greater than about 0.6  as indicated by the solid blue curve. Thus we find that in NMUED a significant hadronic branching fraction is necessary to explain the positron excess which cannot be reconciled with PAMELA antiproton data at the $3\sigma$ level. To further visualize this conflict, in the right panel of Fig.~\ref{fig:fitresults}, we display the range of antiproton fluxes (red shaded region) that corresponds to the $3\sigma$ fit to AMS-02 for a 1~TeV KK photon. We see that even the lowest $\text{BR}_{H}$ overproduces antiprotons by a large margin.

Even if one considers non-power-law $e^\pm$ backgrounds, which permit greater shape variation in the signal $e^\pm$ spectrum to fit AMS-02, the high energy $e^\pm$ spectrum from dark matter will remain almost unchanged to accommodate the positron excess. As a result the maximum $\text{BR}_{H}/\text{BR}_{L}$ bound will also stay the same. 


\begin{figure*}[t]
\centerline{ 
\includegraphics[width=0.4\textwidth]{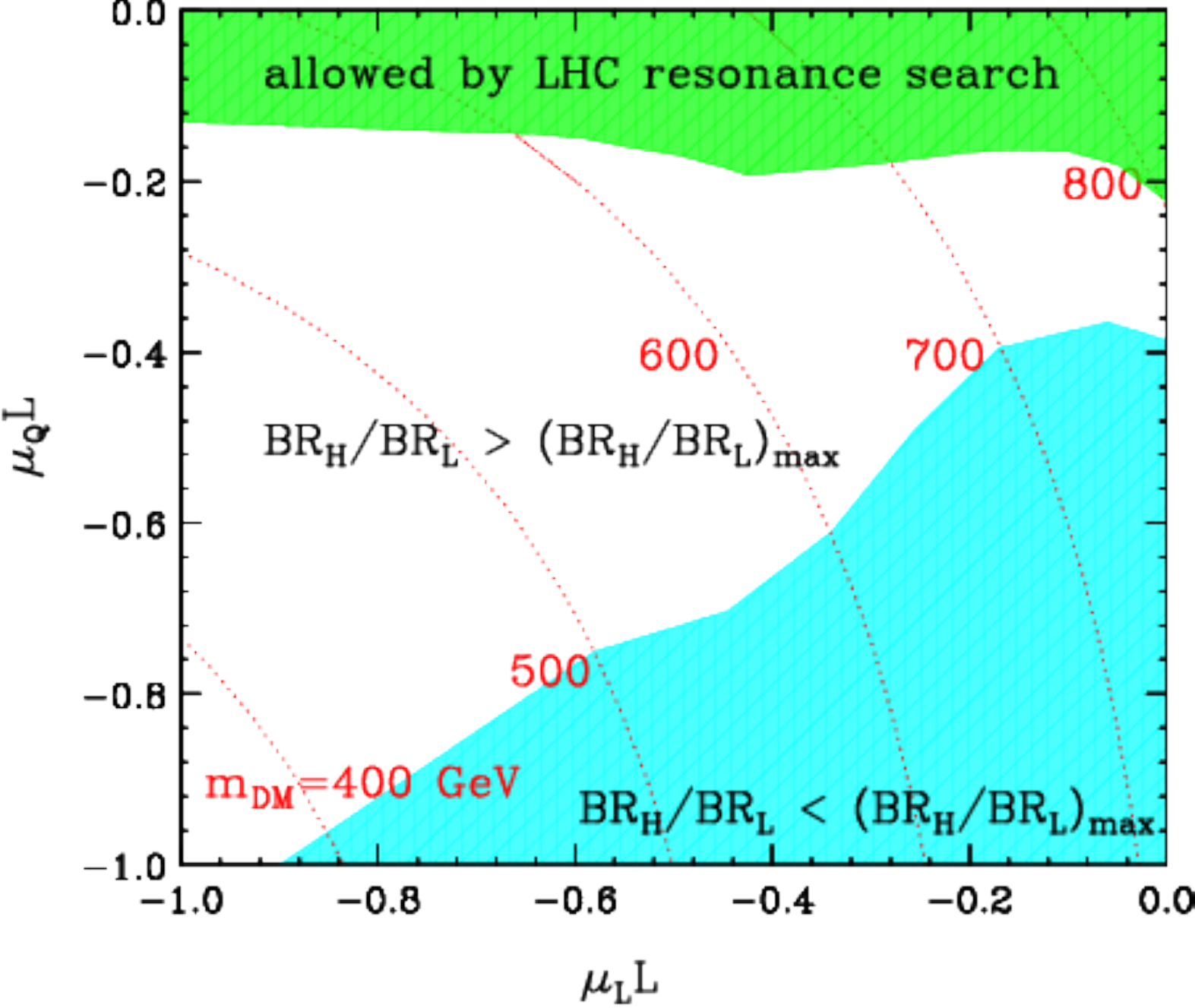}
\hspace{0.2cm}
\includegraphics[width=0.4\textwidth]{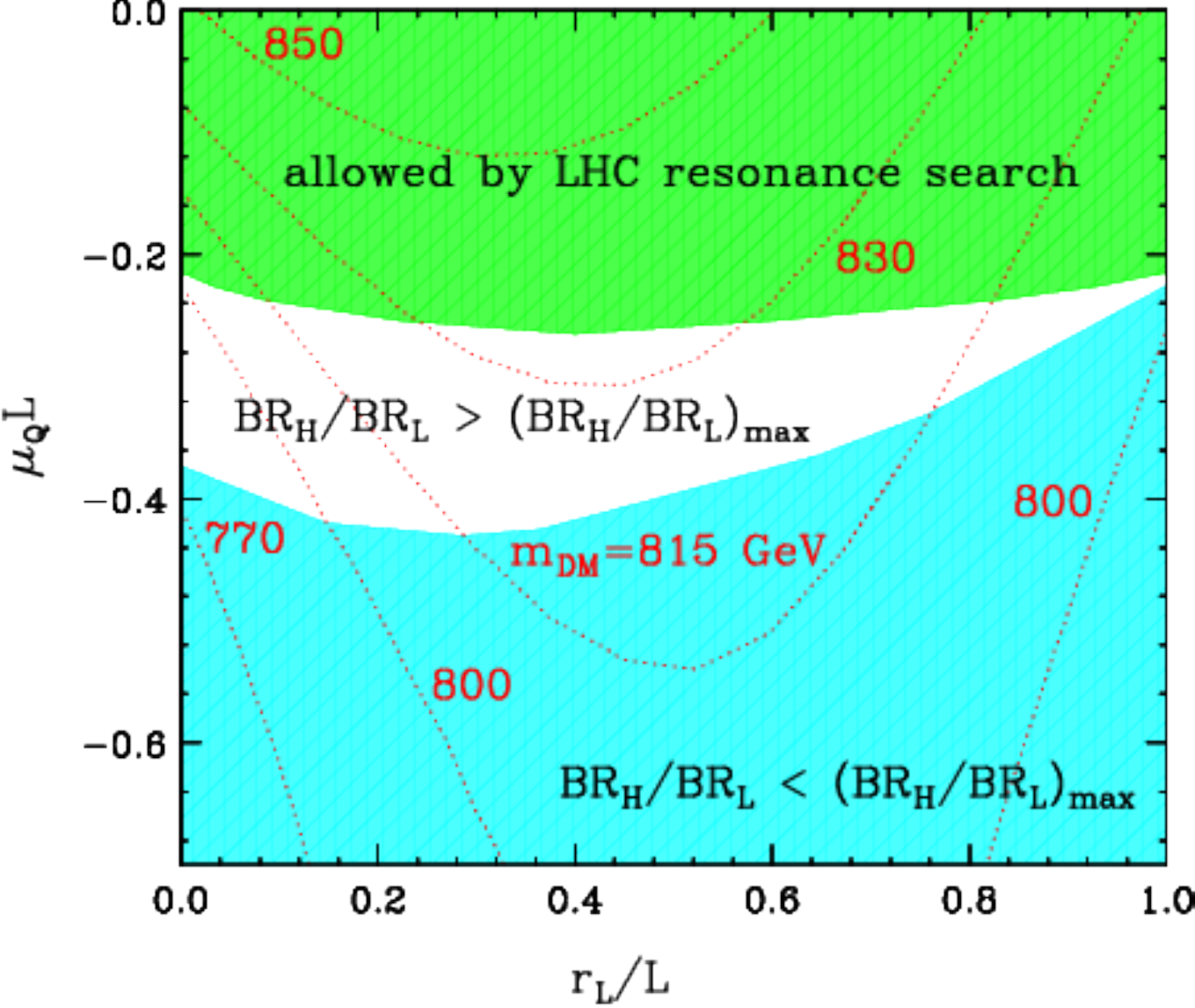}   }
\caption{
Iso-$m_{DM}$ contours in Scenario I, with $r_L=r_Q=0$ and $\mu_L\neq \mu_Q$, (left) and Scenario II, with $\mu_{L} =r_Q=0$, (right) that give $\Omega_{DM} h^2 = 0.11$. The green shaded regions are allowed by resonance searches at the LHC, and the blue shaded regions are compatible with antiproton data.}
\label{fig:nonuniversal}
\end{figure*}

We now check if LHC data supports our conclusion in two special cases that permit different dark matter couplings to leptons and quarks. For completeness we also evaluate the thermal relic abundance of KK photons following the procedure and notation of Refs.~\cite{Huang:2012kz,Flacke:2013pla}.
We take the KK masses to be independent of each other and accordingly rescale the $f_1$-$f_0$-$V_1$ couplings
(where the subscript denotes the KK mode number).
The $V_1$-$H_1$-$H_0$ and $V_1$-$V_1$-$H_0$-$H_0$ couplings remain unchanged due to orthogonality relations. 
We work in the large KK mass limit so that SM particle masses can be neglected. With $\mu, r \neq 0$,
there is a large mass gap between the KK photon and the next lightest KK mode, which renders coannihilation ineffective. Resonant annihilation via KK 2-modes does not occur either since the KK photon masses are not multiples of $R^{-1}$.

In {\it Scenario I}, with $r_L=r_Q=0$ and $\mu_L\neq \mu_Q$,  there are three free parameters, $\mu_L$, $\mu_Q$ and $R^{-1}$.  
Without boundary terms, all the KK boson masses are given by $n/R$ and the KK fermions are heavier than the bosons. 
Since $|\mu_L|$ and  $|\mu_Q|$ suppress the annihilation into leptons and quarks respectively, their sizes determine the positron and antiproton signals. 
However, a large $|\mu|$ increases the coupling between level-2 KK gauge bosons and SM fermion pairs, so that limits from dijet and dilepton searches can be important~\cite{Huang:2012kz}. In {\it Scenario II} with $\mu_{L} =r_Q=0$, the free parameters are $\mu_Q$, $r_L$ and $R^{-1}$. 
Since $\mu_L = 0$, dilepton bounds do not apply, but
a strong bound on $\mu_Q$ from the dijet resonance search is expected. 
In both cases, for a given set of parameters, $R^{-1}$ and $m_{DM}$ are fixed by the relic abundance.

In Fig.~\ref{fig:nonuniversal}, we compare our $\text{BR}_{H}/\text{BR}_{L}$ constraint with collider bounds. 
The (0, 0) point corresponds to the MUED case.
The red dotted curves are iso-$m_{DM}$ contours that reproduce the measured relic density $\Omega_{DM}h^2=0.11$.
Oblique corrections restrict $\mu_L L > -1.5$~\cite{Huang:2012kz}. 

The green shaded regions show the parameter space allowed by LHC dijet searches with 20~fb$^{-1}$ of data at 8~TeV~\cite{lhc1}, and in addition for Scenario~I, dilepton searches with 1~fb$^{-1}$ of data at 7~TeV~\cite{lhc2}. Parameters in the blue shaded region are consistent with the antiproton flux. 
Since the green and blue shaded regions do not overlap in the left panel, Scenario~I is incompatible with LHC data.  
The results of Ref.~\cite{Kong:2013xta} for $r_Q=0$ can be applied to Scenario~II.
The current limit, $\mu_Q L > -0.2$~\cite{Kong:2013xta}, is inconsistent with the blue region.

Within our framework of flavor universality, we extend our analysis beyond the two scenarios considered above
by scanning the four dimensional parameter space in the ranges, $-0.2 < \mu_{L} L \,, \mu_Q L  < 0$ and $0 <  r_{L}/L, r_Q/L< 1$ (where $L=\pi R/2$). We fail to find regions of parameter space
that evade the dijet and dilepton searches and that are consistent with the antiproton data.

\label{sec:concluson}

In summary, we investigated the annihilation of Kaluza-Klein dark matter in UED models extended with brane-localized terms and fermion-bulk masses as an explanation of the AMS-02 and PAMELA positron flux anomaly.
By introducing a hadronic bulk mass term, one can easily suppress the antiproton flux and enhance the positron flux.
However, Next-to-Minimal UED with flavor universality 
cannot explain the cosmic $e^+$ excess and stay consistent with the nonobservation of an antiproton excess because a significant hadronic annihilation branching ratio is required for a soft signal $e^\pm$ spectrum to fit AMS-02. Moreover,  a flavor-blind $\mu_Q$ cannot evade LHC constraints.

%


{\it{Acknowledgements}}. 
D.~M. thanks V.~Bindi and H.~Feldman for discussions.
Y.~G. thanks CHEP for its hospitality during his visit to Peking University while this work was in progress. This work was supported in part by the DOE under Grant Nos. DE-FG02-13ER42024 and DE-FG02-12ER41809, and by the Mitchell Institute for Fundamental Physics and Astronomy. 


\end{document}